\documentclass[12pt,a4paper]{article}%
\usepackage{epsfig}
\usepackage{amsmath}
\usepackage{amsfonts}
\usepackage{amssymb}
\usepackage{graphicx}
\usepackage[T1]{fontenc}
\usepackage[utf8]{inputenc}
\usepackage{authblk}%
\providecommand{\U}[1]{\protect\rule{.1in}{.1in}}
\begin{document}

\title{On the invariant method for the time-dependent non-Hermitian Hamiltonians}
\author{{\small B. Khantoul}$^{a}$\thanks{E-mail: b.khantoul@univ-jijel.dz},
{\small A. Bounames}$^{a}$\thanks{E-mail: bounames@univ-jijel.dz} {\small and}
{\small M. Maamache}$^{b}$\thanks{E-mail: maamache\_m@yahoo.fr }\\$^{(a)}${\small Theoretical Physics Laboratory,} {\small Department of
Physics,}\\{\small University of Jijel, BP 98 \ Ouled Aissa, 18000 Jijel, Algeria.}\\$^{(b)}${\small Laboratoire de Physique Quantique et Syst\`{e}mes Dynamiques,}\\{\small Facult\'{e} des Sciences, Universit\'{e} Ferhat Abbas S\'{e}tif 1,
S\'{e}tif 19000, Algeria}\textit{.} }
\date{}
\maketitle

\begin{abstract}
We propose a scheme to deal with certain time-dependent non-Hermitian
Hamiltonian operators $H(t)$ that generate a real phase in their
time-evolution. This involves the use of invariant operators $I_{PH}\left(
t\right)  $ that are pseudo-Hermitian with respect to the time-dependent
metric operator, which implies that the dynamics is governed by unitary time
evolution. Furthermore, $H(t)$ is generally not quasi-Hermitian and does not
define an observable of the system but $I_{PH}\left(  t\right)  $ obeys a
quasi-hermiticity transformation as in the completely time-independent
Hamiltonian systems case. The harmonic oscillator with a time-dependent
frequency under the action of a complex time-dependent linear potential is
considered as an illustrative example.

PACS: 03.65.Ca, 03.65.-w

\end{abstract}

\section{Introduction}

In Quantum Mechanics, one of the fundamental requirements is that the
Hamiltonian should be Hermitian. Imposing $H^{\dag}=H$ ensures that the
eigenvalue spectrum is real, the inner products of state vectors in Hilbert
space have a positive norm and that the time evolution operator is unitary.
However, it has been found that not only Hermitian Hamiltonians satisfy these
conditions. Specifically, Bender has shown that a non-Hermitian Hamiltonian
which is invariant under $PT$-symmetry satisfies all physical axioms of
quantum theory \cite{Carl1,Carl2,Carl3,Carl4,Carl5}. Parity $P$ has the effect
of changing the sign of the momentum operator $p$ and the position operator
$x$. The anti-linear operator $T$ has the effect of changing the sign of the
momentum operator $p$ and the pure imaginary complex number $i$. The reality
of spectrum was attributed to an unbroken $PT$-symmetry of $H$.

The generalisation of the $PT$-symmetry concept (i.e. systems with real
spectra) to pseudo-Hermiticity was formulated by Mostafazadeh
\cite{most1,most3,most4}: all Hamiltonian $H$ with a real spectrum is
pseudo-Hermitian if%
\begin{equation}
H^{\dag}=\eta H\eta^{-1},\label{1}%
\end{equation}
where the operator $\eta=\rho^{\dag}\rho$ ($\rho$ is a bounded linear
invertible operator, with bounded inverse) being linear, Hermitian, invertible
on the vector space spanned by the eigenstates $\left\vert \phi_{n}%
^{H}\right\rangle $ of $H$. We note that Eq. $\left(  \ref{1}\right)  $ \ is
equivalent to the requirement that $H$ is Hermitian with respect to the inner
product $\left\langle .,.\right\rangle _{\eta}=\left\langle .\left\vert
\eta\right\vert .\right\rangle $ defined as%
\begin{equation}
\langle\phi_{m}^{H}\left\vert \phi_{n}^{H}\right\rangle _{\eta}=\langle
\phi_{m}^{H}|\eta\left\vert \phi_{n}^{H}\right\rangle =\delta_{mn}.
\end{equation}
In particular, the formalism developed by Mostafazadeh, building on earlier
work by Scholtz et al \cite{Scholz}, showed that the Hamiltonian $H$ is
related by a similarity transformation to an equivalent Hermitian Hamiltonian
$h$ by%
\begin{equation}
h=\rho H\rho^{-1},
\end{equation}
the Hermitian Hamiltonian $h$ is equivalent to $H$ in that it has the same
eigenvalue spectrum. Thus, although the eigenvalue spectra of $h$ and $H$ are
identical, relations between their eigenvectors will differ
\begin{equation}
\left\vert \psi_{n}^{h}\right\rangle =\rho\left\vert \phi_{n}^{H}\right\rangle
.
\end{equation}

All these efforts have been devoted to study time-independent non-Hermitian
systems. In contrast, time-dependent non-Hermitian systems are far less well
investigated
\cite{znojil1,most5,most6,susa,fring1,fring2,wang1,wang2,mus,fring4,fring3}
and it appears that so far no consensus has been reached about a number of
central issues. Unexpectedly, a number of conceptual difficulties have been
encountered. Serious problems have arisen, first of all, in connection with
the probabilistic and unitary-evolution interpretation of the generalized
models. The treatment for systems with time-dependent non-Hermitian
Hamiltonians with time-dependent metric operators is still controversially
discussed and was the center of an interesting debate between Mostafazadeh and
Znojil \cite{znojil1,most5,most6}.

In conventional quantum mechanics, the spectral problem for a Hamiltonian or
energy operator is approached by the stationary Schr\"{o}dinger equation. The
general equation of motion is given by the time-dependent Schr\"{o}dinger
equation that describes how a quantum system evolves with time. In this work,
we consider the most general non-Hermitian time-dependent Hamiltonian $H(t)$
and its associated time-dependent metric operator $\eta(t)$.

The main assumption to be made is that the two time-dependent Schr\"{o}dinger
equations still holds%
\begin{equation}
H(t)\left\vert \Phi^{H}(t)\right\rangle =i\hbar\partial_{t}\left\vert \Phi
^{H}(t)\right\rangle ,\label{shrod2}%
\end{equation}%
\begin{equation}
h\left(  t\right)  \left\vert \Psi^{h}(t)\right\rangle =i\hbar\partial
_{t}\left\vert \Psi^{h}(t)\right\rangle ,\label{PSCH}%
\end{equation}
both Hamiltonians involved are explicitly time dependent, with $H(t)$ being to
be non-Hermitian whereas $h(t)$ is taken Hermitian, i.e., $H(t)\neq H^{\dag
}(t)$ and $h(t)=h^{\dag}(t)$. Next, we assume that the two solutions
$\left\vert \Phi^{H}(t)\right\rangle $ and $\left\vert \Psi^{h}%
(t)\right\rangle $ of Eqs. $\left(  \ref{shrod2}\right)  -\left(
\ref{PSCH}\right)  $ are related by a time-dependent invertible operator
$\rho\left(  t\right)  $ as
\begin{equation}
\left\vert \Psi^{h}(t)\right\rangle =\rho\left(  t\right)  \left\vert \Phi
^{H}(t)\right\rangle ,\label{vect}%
\end{equation}
it then follows immediately by direct substitution of $\left(  \ref{vect}%
\right)  $ into Eqs. $\left(  \ref{shrod2}\right)  $ and $\left(
\ref{PSCH}\right)  $ that the two Hamiltonians are allied to each other as
\begin{equation}
h\left(  t\right)  =\rho\left(  t\right)  H\left(  t\right)  \rho^{-1}\left(
t\right)  +i\hbar\dot{\rho}\left(  t\right)  \rho^{-1}\left(  t\right)
,\label{quasi}%
\end{equation}
The key feature in this equation is the fact that $H(t)$ is no longer
quasi-Hermitian, i.e. related to $h(t)$ by means of a similarity
transformation, due to the presence of the last term. Thus $H(t)$ is not a
self-adjoined operator and therefore not observable
\cite{znojil1,fring4,fring3}. From the relation $\left(  \ref{quasi}\right)  $
and using the Hermiticity of $h(t),$ we deduce a relation between $H(t)$ and
its Hermitian conjugate $H^{\dag}(t)$
\begin{equation}
H^{\dag}\left(  t\right)  =\eta\left(  t\right)  H\left(  t\right)  \eta
^{-1}\left(  t\right)  +i\hbar\dot{\eta}\left(  t\right)  \eta^{-1}\left(
t\right)  ,\label{PHH1}%
\end{equation}
the relation $\left(  \ref{PHH1}\right)  $ between the Hamiltonian $H\left(
t\right)  $ and its Hermitian conjugate $H^{\dag}\left(  t\right)  $
generalizes the well known standard quasi-Hermiticity relation $\left(
\ref{1}\right)  $ in the context time-independent non-Hermitian quantum
mechanics \cite{znojil1,fring4,fring3}. Quasi-Hermitian operators are very
special class of pseudo-Hermitian operators. Their importance in physics was
emphasized by Scholtz et al in \cite{Scholz}.

In conventional pseudo-Hermitian (or $PT$-symmetric) theory, when the spectrum
of a non-Hermitian Hamiltonian is purely real the Hamiltonian operator
determines this spectrum through the stationary Schr\"{o}dinger equation, and
when a non-Hermitian Hamiltonian is time-dependent we show that the phases
obtained during the evolution fit the bill.

This work proceeds to investigate in detail the main frames of time-dependent
non-Hermitian systems and goes on to examine how the reality of their phases
can be established. Finally, the original contribution is based on the
definition of pseudo-Hermitian invariant operators, demonstrating a method to
calculate how a quantum system evolves in time with a real phase.

To further elaborate our theoretical proposal, we revisit in Section 2 the
Lewis and Riesenfeld invariant theory problem for an Hermitian harmonic
oscillator systems \cite{Lewis} and we investigate a proper mapping between
conventional invariant theory and pseudo-invariant theory. In Section 3, we
illustrate our time-dependent pseudo-invariant theory by adopting a simple
example: a harmonic oscillator with a time-dependent frequency under the
action of a time-dependent imaginary linear potential.

\section{Invariant operator method}

Here we discuss the advantages of using Lewis and Riesenfeld invariant
operator method in explicitly time-dependent quantum systems by giving a brief
review \cite{Lewis}. We consider a system whose Hamiltonian $h(t)$ is
Hermitian and explicitly time dependent. A Hermitian operator $I_{h}\left(
t\right)  $ is called an invariant for the system if it satisfies
\begin{equation}
\frac{dI_{h}(t)}{dt}=\frac{\partial I_{h}(t)}{\partial t}-\frac{i}{\hbar
}\left[  I_{h}\left(  t\right)  ,h\left(  t\right)  \right]  =0. \label{LR}%
\end{equation}

The eigenvalue equation of $I_{h}\left(  t\right)  $ can be written as
\begin{equation}
I_{h}\left(  t\right)  \left\vert \psi_{n}^{h}(t)\right\rangle =\lambda
_{n}\left\vert \psi_{n}^{h}(t)\right\rangle . \label{Eveq}%
\end{equation}
With the help of Eq. $\left(  \ref{LR}\right)  $, it is easy to show that the
real eigenvalues $\lambda_{n}$ are time-independent. The Schr\"{o}dinger
equation $\left(  \ref{PSCH}\right)  $ for the system has particular solutions
$\left\vert \Psi_{n}^{h}(t)\right\rangle $ different from $\left\vert \psi
_{n}^{h}(t)\right\rangle $ in Eq. $\left(  \ref{Eveq}\right)  $ only by a
phase factor $e^{i\varepsilon_{n}(t)}$ where the phase $\varepsilon_{n}(t)$ is
given by
\begin{equation}
\hbar\frac{d}{dt}\varepsilon_{n}(t)=\left\langle \psi_{n}^{h}(t)\right\vert
i\hbar\frac{\partial}{\partial t}-h\left(  t\right)  \left\vert \psi_{n}%
^{h}(t)\right\rangle . \label{pha}%
\end{equation}

The first special physical system to which Lewis and Riesenfeld \cite{Lewis}
have applied their general result is that of a time-dependent harmonic
oscillator for which the frequency parameter is allowed to vary with time
\begin{equation}
h_{osc}\left(  t\right)  =\frac{p^{2}}{2m}+\frac{1}{2}m\omega^{2}(t)x^{2}.
\label{h_osc}%
\end{equation}
They derive an exact invariant for this system by means of the equation
$\left(  \ref{LR}\right)  ,$ that is
\begin{equation}
I_{h}^{osc}(t)=\sigma^{2}\left(  t\right)  p^{2}-m\sigma\left(  t\right)
\dot{\sigma}\left(  t\right)  \left[  px+xp\right]  +\frac{1}{\sigma
^{2}\left(  t\right)  }\left[  1+m^{2}\sigma^{2}\left(  t\right)  \dot{\sigma
}^{2}\left(  t\right)  \right]  x^{2}, \label{INV_h}%
\end{equation}
where $\sigma\left(  t\right)  $ satisfies the non-linear auxiliary equation%
\begin{equation}
\ddot{\sigma}\left(  t\right)  +\sigma\left(  t\right)  \omega^{2}(t)=\frac
{1}{m^{2}\sigma^{3}\left(  t\right)  }. \label{auxi}%
\end{equation}
Then, the eigenstates and eigenvalues of this invariant are
\cite{Lewis,pedrosa}%
\begin{equation}
\psi_{n}^{I_{h}^{osc}}\left(  x,t\right)  =\left[  \frac{1}{n!2^{n}\sigma
\sqrt{\pi\hbar}}\right]  ^{\frac{1}{2}}\exp\left[  \frac{im}{2\hbar}\left(
\frac{\dot{\sigma}}{\sigma}+\frac{i}{m\sigma^{2}}\right)  x^{2}\right]
H_{n}\left[  \left(  \frac{1}{\hbar}\right)  ^{\frac{1}{2}}(\frac{x}{\sigma
})\right]  ,\text{ \ \ \ \ }\lambda_{n}=\hbar(n+\frac{1}{2}),
\label{eingfuc_h}%
\end{equation}
where $H_{n}$ is the usual Hermite polynomial of order $n$, and the
appropriate time-dependent phase factor that make the eigenstates solutions of
the Schr\"{o}dinger equation is
\begin{equation}
\epsilon_{n}\left(  t\right)  =-\left(  n+\frac{1}{2}\right)  \int_{0}%
^{t}\frac{1}{m\sigma^{2}\left(  t^{\prime}\right)  }dt^{\prime}. \label{C0}%
\end{equation}

Now we proceed introducing and analyzing the spectral properties of
pseudo-Hermitian invariant operator $I_{PH}\left(  t\right)  $. Particular
attention is given to the special subset of quasi-Hermitian operators. We
start by considering a non-Hermitian quantum mechanics in its most general
form by studying time-dependent Hamiltonian operators $H(t)$ and also
time-dependent metric operator $\eta\left(  t\right)  =\rho^{\dag}\left(
t\right)  \rho\left(  t\right)  $ associated with $H(t)$. In the study of the
time evolution problem, let us admit that a time-dependence occurs in all the
operators. A non-Hermitian operator $I_{PH}\left(  t\right)  $ is said to be a
pseudo-Hermitian operator if it satisfies%
\begin{equation}
I_{PH}^{\dag}\left(  t\right)  =\eta(t)I_{PH}\left(  t\right)  \eta
^{-1}(t)\text{ }\Leftrightarrow I_{h}(t)=\rho(t)I_{PH}(t)\rho^{-1}%
(t)=I_{h}^{\dag}(t).\label{quas}%
\end{equation}

The virtue of such a conjugate pair $I_{h}(t)$ and $I_{PH}(t)$ is that they
possess an identical eigenvalue spectrum because the invariants lie in the
same similarity class. The reality of the spectrum is guaranteed, since one of
the invariants involved, i.e. $I_{h}(t),$ is Hermitian. It means that any
self-adjoined invariant operator $I_{h}(t)$, i.e. observable, in the Hermitian
system possesses an invariant counterpart $I_{PH}(t)$ in the non-Hermitian
system given by $I_{PH}(t)$ $=\rho^{-1}(t)I_{h}(t)\rho(t)$ in complete analogy
to the time-independent scenario for any self-adjoint operator.

The corresponding eigenvalue equations are then simply
\begin{equation}
I_{h}\left(  t\right)  \left\vert \psi_{n}^{h}(t)\right\rangle =\lambda
_{n}\left\vert \psi_{n}^{h}(t)\right\rangle ,\text{ and \ }I_{PH}\left(
t\right)  \left\vert \phi_{n}^{PH}(t)\right\rangle =\lambda_{n}\left\vert
\phi_{n}^{PH}(t)\right\rangle , \label{I_PH}%
\end{equation}
where the eigenfunctions $\left\vert \psi_{n}^{h}(t)\right\rangle $ and
$\left\vert \phi_{n}^{PH}(t)\right\rangle $ are related as
\begin{equation}
\left\vert \psi_{n}^{h}(t)\right\rangle =\rho(t)\left\vert \phi_{n}%
^{PH}(t)\right\rangle .\text{ }%
\end{equation}
The inner product for the eigenfunctions $\left\vert \phi_{n}^{PH}%
(t)\right\rangle $ related to the pseudo-Hermitian invariant $I_{PH}(t)$
satisfies
\begin{equation}
\langle\phi_{m}^{PH}(t)\left\vert \phi_{n}^{PH}(t)\right\rangle _{\eta
}=\langle\phi_{m}^{PH}(t)|\eta\left\vert \phi_{n}^{PH}(t)\right\rangle
=\delta_{mn}. \label{11}%
\end{equation}

It is easy to verify by direct computation that the $I_{PH}\left(  t\right)  $
defined by Eq. $\left(  \ref{quas}\right)  $ satisfies
\begin{equation}
\frac{\partial I_{PH}(t)}{\partial t}=\frac{i}{\hbar}\left[  I_{PH}%
(t),H(t)\right]  , \label{NINV}%
\end{equation}
with a non-Hermitian Hamiltonian $H(t),$ which govern the time evolution of
Schr\"{o}dinger equation, given by Eq. $\left(  \ref{shrod2}\right)  $. The
eigenstates and eigenvalues of the invariant operator $I_{PH}(t)$ may be found
by the same technique completely analogous to the method introduced above for
the Hermitian case. It is, of course, natural to calculate the solution of the
non-Hermitian time-dependent Schr\"{o}dinger equation $\left(  \ref{shrod2}%
\right)  $ as in the time-dependent Hermitian case.

The Schr\"{o}dinger equation $\left(  \ref{shrod2}\right)  $ for the system
has particular solutions $\left\vert \Phi_{n}^{H}(t)\right\rangle $ different
from $\left\vert \phi_{n}^{PH}(t)\right\rangle $ in Eq. $\left(
\ref{I_PH}\right)  $ only by a phase factor $e^{i\varepsilon_{n}^{PH}(t)}$
where the phase $\varepsilon_{n}^{PH}(t)$ is given by%
\begin{equation}
\hbar\frac{d}{dt}\varepsilon_{n}^{PH}(t)=\left\langle \phi_{n}^{PH}%
(t)\right\vert \eta(t)\left[  i\hbar\frac{\partial}{\partial t}-H\left(
t\right)  \right]  \left\vert \phi_{n}^{PH}(t)\right\rangle . \label{phase1}%
\end{equation}
In Eq. $\left(  \ref{phase1}\right)  $, the first term is parallel to a
familiar non adiabatic geometrical phase and the second term represents the
dynamical effect. It is the sum of these two terms that can ensure a real
total phase $\varepsilon_{n}^{PH}(t).$

In the end, it is important to note that the Schrodinger equation for
explicitly time-dependent Hamiltonian cannot be written in the form of an
eigenvalue equation and therefore, in this case, nothing can be said about the
spectrum of the Hamiltonian, and consequently, we are interested in its mean
value. However, the invariant operator satisfies an eigenvalue equation with a
real time-independent spectrum.

\section{ Time-dependent Harmonic oscillator with a complex time-dependent
potential}

As an application, we study an oscillator with time-dependent frequency under
the action of a time-dependent imaginary linear potential, and we compare our
results with those obtained in Ref. \cite{susa}%
\begin{equation}
H(t)=\frac{p^{2}}{2m}+\frac{1}{2}m\omega^{2}(t)x^{2}+i\lambda(t)x, \label{HH}%
\end{equation}
where $\lambda(t)$ is a real time-dependent function.

Let us take the metric operator in the form%
\begin{equation}
\eta^{-1}\left(  t\right)  =\exp\frac{1}{\hbar}[\alpha\left(  t\right)
p+\beta\left(  t\right)  x],\text{ }%
\end{equation}
where $\alpha\left(  t\right)  $ and $\beta\left(  t\right)  $ are unknown
real time-dependent functions$.$

Using Eq. $\left(  \ref{PHH1}\right)  $, we obtain
\begin{equation}
\left.
\begin{array}
[c]{c}%
\dot{\alpha}\left(  t\right)  +\frac{\beta\left(  t\right)  }{m}=0,\\
m\omega^{2}\left(  t\right)  \alpha\left(  t\right)  +2\lambda\left(
t\right)  -\dot{\beta}\left(  t\right)  =0.
\end{array}
\right.
\end{equation}
From the first equation $\beta\left(  t\right)  =-m\dot{\alpha}\left(
t\right)  $, the second equation can be reduced to
\begin{equation}
m\overset{\cdot\cdot}{\alpha}\left(  t\right)  +m\omega^{2}\left(  t\right)
\alpha\left(  t\right)  +2\lambda\left(  t\right)  =0.\label{diff}%
\end{equation}
Then the time-dependents metric operators $\eta\left(  t\right)  $ is given
by
\[
\eta\left(  t\right)  =\exp\left[  -\frac{\alpha\left(  t\right)  }{\hbar
}p+\frac{m\dot{\alpha}\left(  t\right)  }{\hbar}x\right]  ,
\]
using the relation $\eta\left(  t\right)  =\rho^{\dag}\left(  t\right)
\rho\left(  t\right)  $ where $\rho\left(  t\right)  $ is not unique and can
be taken as a real operator
\begin{equation}
\rho(t)=\exp\left[  -\frac{\alpha\left(  t\right)  }{2\hbar}p+\frac
{m\dot{\alpha}\left(  t\right)  }{2\hbar}x\right]  .\label{rho}%
\end{equation}
It can easily be shown that under the transformation $\rho(t)$ defined in Eq.
$\left(  \ref{rho}\right)  ,$ the coordinate and momentum operators change
according to%
\begin{equation}
\rho^{-1}(t)x\rho(t)=x-i\frac{\alpha}{2},\text{ \ \ \ \ \ }\rho^{-1}%
(t)p\rho(t)=p-i\frac{m\dot{\alpha}}{2}.\label{rho2}%
\end{equation}

An important property of the transformation $\rho^{-1}(t)$, the action of
which on a wave function in the $x$-representation reads%
\begin{equation}
\rho^{-1}G(x)=\exp\left[  i\frac{\hbar}{8}m\alpha\dot{\alpha}\right]
\exp\left[  -\frac{m\dot{\alpha}\left(  t\right)  }{2}x\right]  G(x-i\frac
{\alpha}{2}). \label{rho1}%
\end{equation}

To affect the evaluation of the phase $\left(  \ref{phase1}\right)  $, we need
to calculate the diagonal matrix elements of the operators $H\left(  t\right)
$ and $i\hbar\frac{\partial}{\partial t}$. That is%
\begin{align}
\left\langle \phi_{n}^{PH}(t)\right\vert \eta(t)\left[  i\hbar\frac{\partial
}{\partial t}-H\left(  t\right)  \right]  \left\vert \phi_{n}^{PH}%
(t)\right\rangle  &  =\left\langle \phi_{n}^{PH}(t)\right\vert \rho^{\dag
}(t)\rho(t)\left[  i\hbar\frac{\partial}{\partial t}-H\left(  t\right)
\right]  \rho^{-1}(t)\rho(t)\left\vert \phi_{n}^{PH}(t)\right\rangle
\nonumber\\
&  =\left\langle \phi_{n}^{PH}(t)\right\vert \rho^{\dag}(t)\left(  i\hbar
\frac{\partial}{\partial t}-\rho(t)H\left(  t\right)  \rho^{-1}(t)\right.
\nonumber\\
&  \left.  -i\hbar\rho\left(  t\right)  \frac{\partial}{\partial t}\rho
^{-1}\left(  t\right)  \right)  \rho(t)\left\vert \phi_{n}^{PH}%
(t)\right\rangle \label{C1}%
\end{align}
using Eq. $\left(  \ref{rho2}\right)  $ we express $\rho H\left(  t\right)
\rho^{-1}$ as%

\begin{align}
\rho(t)H\left(  t\right)  \rho^{-1}(t) &  =\frac{p^{2}}{2m}+\frac{1}{2}%
m\omega^{2}\left(  t\right)  x^{2}+i\left(  \lambda+\frac{m\alpha\omega
^{2}\left(  t\right)  }{2}\right)  x\nonumber\\
&  +i\frac{\overset{\cdot}{\alpha}}{2}p-\left(  m\frac{\overset{\cdot}{\alpha
}^{2}}{8}+\frac{m\alpha^{2}\omega^{2}\left(  t\right)  }{8}+\frac
{\alpha\lambda}{2}\right)  ,\label{C2}%
\end{align}
and taking the partial time derivative of $\rho^{-1}$ we obtain the
appropriate product
\begin{equation}
i\hbar\rho\left(  t\right)  \dot{\rho}^{-1}\left(  t\right)  =i\frac
{\dot{\alpha}}{2}p-im\frac{\ddot{\alpha}}{2}x-\frac{m}{8}\left(  \dot{\alpha
}^{2}-\alpha\ddot{\alpha}\right)  .\label{C3}%
\end{equation}
The diagonal matrix elements of the operator $\left(  i\hbar\frac{\partial
}{\partial t}-H\right)  $ can be simplified by using Eq. $\left(
\ref{diff}\right)  $
\begin{align}
\left\langle \phi_{n}^{PH}(t)\right\vert \eta(t)\left[  i\hbar\frac{\partial
}{\partial t}-H\left(  t\right)  \right]  \left\vert \phi_{n}^{PH}%
(t)\right\rangle  &  =\left\langle \phi_{n}^{PH}(t)\right\vert \rho^{\dag
}(t)\left[  -\left(  \frac{p^{2}}{2m}+\frac{1}{2}m\omega^{2}\left(  t\right)
x^{2}-\frac{\alpha\lambda}{4}\right)  \right.  \nonumber\\
&  \left.  +i\hbar\frac{\partial}{\partial t}\right]  \rho(t)\left\vert
\phi_{n}^{PH}(t)\right\rangle .\label{C4}%
\end{align}
The key feature in this equation is the fact that the eigenstates of
$I_{PH}\left(  t\right)  $ are related to those of the Hermitian Invariant
$I_{h}(t)$ by $\left\vert \phi_{n}^{PH}(t)\right\rangle =\rho^{-1}%
(t)\left\vert \psi_{n}^{h}(t)\right\rangle $ , because $I_{PH}\left(
t\right)  $ is quasi-Hermitian, i.e. related to $I_{h}(t)$ by means of the
similarity transformation $\left(  \ref{quas}\right)  .$ Therefore, the
time-dependent c-number $\hbar\alpha\left(  t\right)  \lambda(t)/4$ can be
removed by a time-dependent unitary transformation namely $\exp\left[
i\int_{0}^{t}\frac{\alpha\left(  t^{\prime}\right)  \lambda(t^{\prime})}%
{4}dt^{\prime}\right]  $ to give the time derivative of the phase
\begin{align}
\left\langle \phi_{n}^{PH}(t)\right\vert \eta(t)\left[  i\hbar\frac{\partial
}{\partial t}-H\left(  t\right)  \right]  \left\vert \phi_{n}^{PH}%
(t)\right\rangle  &  =\hbar\frac{d\epsilon_{n}\left(  t\right)  }%
{dt}\nonumber\\
&  =\left\langle \psi_{n}^{h}(t)\right\vert i\hbar\frac{\partial}{\partial
t}-\left(  \frac{p^{2}}{2m}+\frac{1}{2}m\omega^{2}\left(  t\right)
x^{2}\right)  \left\vert \psi_{n}^{h}(t)\right\rangle .\label{C5}%
\end{align}
We recognize the phase associated to the time-dependent one-dimensional
harmonic oscillator system whose Hamiltonian operator is given in Section 2 by
Eq. $\left(  \ref{h_osc}\right)  $ and therefore the associated Hermitian
invariant operator $I_{h}^{osc}$, its eigenstates $\psi_{n}^{I_{h}^{osc}%
}\left(  x,t\right)  $ and the phase $\epsilon_{n}$ are given by the equations
$\left(  \ref{INV_h}\right)  ,\left(  \ref{eingfuc_h}\right)  $ and $\left(
\ref{C0}\right)  $ respectively.

By using the quasi-Hermiticity equation $\left(  \ref{quas}\right)  $, the
pseudo-Hermitian invariant associated to the non-Hermitian Hamiltonian
$H\left(  t\right)  $ can be easily obtained
\begin{align}
I_{PH}  &  =\sigma^{2}\left(  t\right)  \left(  p-i\frac{m\dot{\alpha}}%
{2}\right)  ^{2}-m\sigma\left(  t\right)  \dot{\sigma}\left(  t\right)
\left[  \left(  p-i\frac{m\dot{\alpha}}{2}\right)  \left(  x-i\frac{\alpha}%
{2}\right)  +\left(  x-i\frac{\alpha}{2}\right)  \left(  p-i\frac{m\dot
{\alpha}}{2}\right)  \right] \nonumber\\
&  \left.  +\frac{1}{\sigma^{2}\left(  t\right)  }\left[  1+m^{2}\sigma
^{2}\left(  t\right)  \dot{\sigma}^{2}\left(  t\right)  \right]  \left(
x-i\frac{\alpha}{2}\right)  ^{2}\right.  . \label{pseudo inv}%
\end{align}
Thus, the phase of evolved state $\left\vert \Phi_{n}^{PH}(t)\right\rangle $
are real and can be obtained with the help of Eqs. $\left(  \ref{C0}\right)  $
and $\left(  \ref{C5}\right)  $
\begin{equation}
\epsilon_{n}^{PH}\left(  t\right)  =-\int_{0}^{t}\left[  \left(  n+\frac{1}%
{2}\right)  \frac{1}{m\sigma^{2}\left(  t^{\prime}\right)  }-\frac
{\lambda\left(  t^{\prime}\right)  \alpha\left(  t^{\prime}\right)  }{4\hbar
}\right]  dt^{\prime}. \label{PHASE}%
\end{equation}
However, the general solution for the time-dependent Schr\"{o}dinger equation
of the non-Hermitian Hamiltonian $\left(  \ref{HH}\right)  $ is given by%
\begin{equation}
\Phi_{n}^{H}\left(  x,t\right)  =\exp\left[  i\epsilon_{n}^{PH}\left(
t\right)  \right]  \phi_{n}^{PH}\left(  x,t\right)  =\exp\left[  -i\int
_{0}^{t}\left[  \left(  n+\frac{1}{2}\right)  \frac{1}{m\sigma^{2}\left(
t^{\prime}\right)  }-\frac{\lambda\left(  t^{\prime}\right)  \alpha\left(
t^{\prime}\right)  }{4\hbar}\right]  dt^{\prime}\right]  \phi_{n}^{PH}\left(
x,t\right)  , \label{final}%
\end{equation}
where
\begin{equation}
\phi_{n}^{PH}\left(  x,t\right)  =\rho^{-1}(t)\psi_{n}^{I_{h}^{osc}}%
(x,t)=\exp\left[  i\frac{\hbar}{8}m\alpha\dot{\alpha}\right]  \exp\left[
-\frac{m\dot{\alpha}\left(  t\right)  }{2}x\right]  \psi_{n}^{I_{h}^{osc}%
}\left(  x-i\frac{\alpha}{2},t\right)  , \label{sol}%
\end{equation}
are eigenfunctions of $I_{PH}(t)$ obtained by the inverse tansformation on
eigenfunctions $\left(  \ref{eingfuc_h}\right)  $ \ of $I_{h}^{osc}(t)$.

Before concluding let us make a few remarks about the nature of the solution
in certain special cases. A particular example is a harmonic oscillator with a
time-dependent $PT$-violating linear potential \cite{susa} where the frequency
$\omega=\omega_{0}$ is constant and $\lambda\left(  t\right)  =at.$\ \ Here,
the equations $\left(  \ref{auxi}\right)  $ and $\left(  \ref{diff}\right)  $
for $\sigma(t)$ and\ $\alpha\left(  t\right)  $ can be explicitly solved to
yield
\begin{equation}
\frac{1}{m\sigma^{2}}=\omega_{0},\text{ \ \ \ \ \ \ \ \ \ \ }\alpha\left(
t\right)  =-\frac{2at}{m\omega_{0}^{2}}.
\end{equation}
Then, the phase $\left(  \ref{PHASE}\right)  $ can be determined as
\begin{equation}
\epsilon_{n}\left(  t\right)  =-\left(  n+\frac{1}{2}\right)  \omega
_{0}t-\frac{a^{2}t^{3}}{6\hbar m\omega_{0}^{2}},\label{PH1}%
\end{equation}
and\ our new wave function $\left(  \ref{final}\right)  $ reduces to those
obtained in Ref. \cite{susa}
\begin{align}
\Phi_{n}^{H}\left(  x,t\right)   &  =\exp\left[  i\epsilon_{n}\left(
t\right)  \right]  \exp\left[  i\frac{\hbar a^{2}t}{2m\omega_{0}^{4}}\right]
\exp\left[  \frac{a}{\omega_{0}^{2}}x\right]  \left[  \frac{1}{n!2^{n}}%
\sqrt{\frac{m\omega_{0}}{\pi\hbar}}\right]  ^{\frac{1}{2}}\nonumber\\
&  \exp\left[  -\left(  \frac{m\omega_{0}}{2\hbar}\right)  \left(
x+i\frac{at}{m\omega_{0}^{2}}\right)  ^{2}\right]  H_{n}\left[  \left(
\frac{m\omega_{0}}{\hbar}\right)  ^{\frac{1}{2}}\left(  x+i\frac{at}%
{m\omega_{0}^{2}}\right)  \right]  ,
\end{align}
where the phase functions $\epsilon_{n}\left(  t\right)  $ are given by Eq.
$\left(  \ref{PH1}\right)  $.

\section{Conclusion}

In this work, we studied a class of general explicitly time-dependent
non-Hermitian problems in quantum mechanics, e.g., those with a
\ time-dependent pseudo-Hermitian invariant operator and a time-dependent
metric $\eta(t)$ which have raised a controversy \cite{znojil1,most5,most6}.
Because a non-Hermitian time-dependent Hamiltonian $H(t)$ whose associated
Schr\"{o}dinger equation $\left(  \ref{shrod2}\right)  $ is mapped, by means
of the time-dependent operator $\rho(t)$, into the Schr\"{o}dinger equation
$\left(  \ref{PSCH}\right)  $, where the corresponding wave functions are
transformed as $\left\vert \Phi^{H}(t)\right\rangle =\rho^{-1}(t)\left\vert
\Psi^{h}(t)\right\rangle $ and the Hamiltonians are related by means of the
time-dependent relation $\left(  \ref{quasi}\right)  $. Thus $H(t)$ and $h(t)$
are no longer related by a quasi-hermiticity transformation as in the
completely time-independent case \cite{Dyson} or the time-dependent case with
time-independent metric \cite{fring1,fring2}, but instead their mutual
dependence involves an additional time-dependent term $-i\hbar\rho^{-1}\left(
t\right)  \dot{\rho}\left(  t\right)  .$ The authors of \cite{fring4,fring3}
refer to Eq. $\left(  \ref{quasi}\right)  $ as the time-dependent
quasi-Hermiticity relation and of course the non-Hermitian Hamiltonian $H(t$)
does not belong to the set of observables in this system as it is not related
to $h(t)$ by a similarity transformation. \ In our circumstance, it is evident
that the self-adjoint invariant operator $I_{h}(t)$ associated with the
Hermitian Hamiltonian $h(t)$, i.e., an observable, in the Hermitian system has
an invariant observable counterpart $I_{PH}(t)$ associated with a
non-Hermitian Hamiltonian $H(t)$ in the non-Hermitian system related to each
other as $I_{h}(t)=\rho(t)I_{PH}(t)\rho^{-1}(t)$, since
\begin{align}
\langle\phi_{m}^{H}\left\vert I_{PH}\phi_{n}^{H}\right\rangle _{\eta}  &
=\langle I_{PH}\phi_{m}^{H}\left\vert \phi_{n}^{H}\right\rangle _{\eta
}=\langle\psi_{m}^{h}\left\vert I_{h}\psi_{n}^{h}\right\rangle \nonumber\\
&  =\langle I_{h}\psi_{m}^{h}\left\vert \psi_{n}^{h}\right\rangle =\lambda
_{n}\delta_{mn}%
\end{align}
both invariants $I_{PH}(t)$ and $I_{h}(t)$ possess an identical eigenvalue
spectrum because the invariants lie in the same similarity class. The reality
of the spectrum is guaranteed, since one of the invariants involved, i.e.
$I_{h}(t),$ is Hermitian. We have shown that the evolved state of a
time-dependent non-Hermitian quantum systems acquires a real phase during its
evolution. Therefore, the Lewis and Riesenfeld phase is invariant under the
transformation $\rho(t)$
\begin{align}
\hbar\frac{d}{dt}\varepsilon_{n}^{PH}(t)  &  =\left\langle \phi_{n}%
^{PH}(t)\right\vert \eta(t)\left[  i\hbar\frac{\partial}{\partial t}-H\left(
t\right)  \right]  \left\vert \phi_{n}^{PH}(t)\right\rangle \nonumber\\
&  =\left\langle \psi_{n}^{h}(t)\right\vert \left[  i\hbar\frac{\partial
}{\partial t}-h\left(  t\right)  \right]  \left\vert \psi_{n}^{h}%
(t)\right\rangle =\hbar\frac{d}{dt}\varepsilon_{n}(t)
\end{align}

This is due essentially to the derivation, for a \ pseudo-Hermitian invariant,
of the Liouville equation (\ref{NINV}) which is exactly similar to the
Hermitian case (\ref{LR}) where $I_{h}\left(  t\right)  $ and $h(t)$ are
replaced by $I_{PH}(t)$ and $H(t)$.

Finally, our formalism has been applied to find the solution of the harmonic
oscillator with time dependent frequency under the action of a complex
time-dependent linear potential. At this point, we are able to calculate the
expectation value of the Hamiltonian of the system. For this we calculate the
mean value of the Hamiltonian in a closed form, as usual, using the above
result in Eq.( \ref{C2})
\begin{align}
\langle H\left(  t\right)  \rangle_{\eta}  &  =\left\langle \phi_{n}%
^{PH}(t)\right\vert \eta(t)H\left(  t\right)  \left\vert \phi_{n}%
^{PH}(t)\right\rangle =\left\langle \psi_{n}^{h}(t)\right\vert \rho(t)H\left(
t\right)  \rho^{-1}(t)\left\vert \psi_{n}^{h}(t)\right\rangle \nonumber\\
&  =\frac{\hbar}{2}\left(  n+\frac{1}{2}\right)  \left(  \overset{\cdot
}{\sigma}^{2}+m\omega^{2}\left(  t\right)  \sigma^{2}+\frac{1}{m\sigma^{2}%
}\right)  -\left(  m\frac{\overset{\cdot}{\alpha}^{2}}{8}+\frac{m\alpha
^{2}\omega^{2}\left(  t\right)  }{8}+\frac{\alpha\lambda}{2}\right)  ,
\label{mean}%
\end{align}
where we have used the following mean values%

\[
\left\langle \psi_{n}^{h}(t)\right\vert x\left\vert \psi_{n}^{h}%
(t)\right\rangle =\left\langle \psi_{n}^{h}(t)\right\vert p\left\vert \psi
_{n}^{h}(t)\right\rangle =0
\]
and \cite{Lewis}
\[
\left\langle \psi_{n}^{h}(t)\right\vert \frac{p^{2}}{2m}+\frac{1}{2}%
m\omega^{2}\left(  t\right)  x^{2}\left\vert \psi_{n}^{h}(t)\right\rangle
=\frac{\hbar}{2}\left(  n+\frac{1}{2}\right)  \left(  \overset{\cdot}{\sigma
}^{2}+m\omega^{2}\left(  t\right)  \sigma^{2}+\frac{1}{m\sigma^{2}}\right)  .
\]

For the particular case $\omega=\omega_{0}$ and $\lambda\left(  t\right)
=at,$ Eq $\left(  \ref{mean}\right)  $ is reduced to
\[
\langle H\left(  t\right)  \rangle_{\eta}=\hbar\omega_{0}\left(  n+\frac{1}%
{2}\right)  +\frac{a^{2}t^{2}}{2m\omega_{0}^{2}}-\frac{a^{2}}{2m\omega_{0}%
^{4}},
\]
which coincides with the result obtained in Ref. \cite{susa} $\left(
\text{for the choice }\dot{\eta}=0\text{ in Eq. }\left(  28\right)  \text{ in
\cite{susa}}\right)  .$

Then we have proven that the reality of the time-dependent mean
value\ $\langle H\left(  t\right)  \rangle_{\eta}$ is maintained despite the
non-hermiticity of the Hamiltonian.

\paragraph{Acknowledgments}

Two of the authors (A. B and M. M) would like to thank Professor Andreas Fring
for the interesting discussions on the notion of the time-dependent
quasi-Hermiticity during the PHHQP16 workshop.

\end{document}